\documentclass{llncs}

\usepackage{amssymb}
\usepackage{mathtools}
\usepackage{multicol}
\usepackage{makeidx}

\usepackage[utf8]{inputenc}

\usepackage{cite} 
\newcommand\citep{\cite}
\newcommand\citet{\cite}

\usepackage{hyperref}
\usepackage{xcolor}
\usepackage{tabulary}
\usepackage{enumerate}
\usepackage{tikz}
\usepackage{tikz-qtree}
\usepackage{nameref}

\usepackage[ruled,vlined,inoutnumbered]{algorithm2e}
\SetKwBlock{Proc}{procedure}{end procedure}
\SetKwProg{Fn}{function}{}{end function}
\SetKwInput{KwIn}{In}
\SetKwInput{KwOut}{Out}
\SetKwInput{KwVars}{Vars}
\SetKwBlock{BlockInput}{Input:}{}
\SetKwBlock{BlockAction}{}{end}
\SetKw{True}{$\mathtt{true}$}
\SetKw{False}{$\mathtt{false}$}
\SetKw{Nil}{$\mathtt{nil}$}
\SetKw{Not}{$\mathbf{not}$}

\newcommand{\tuple}[1]{\left\langle #1 \right\rangle}
\newcommand\tup{\tuple}
\newcommand{\logexpr}[1]{[\hspace*{-1.5pt}[\, #1 \,]\hspace*{-1.5pt}]}
\newcommand\wc{{\Large \textunderscore}}

\newcommand{\PRE}{\textsf{PRE}}
\newcommand{\POST}{\textsf{POST}}
\newcommand{\SELX}{\textsf{SELX}}

\newcommand{\Ext}{\textit{\textsc{Ext}}}

\newcommand{\LS}{\textit{\textsc{LS}}}
\newcommand{\RR}{\textit{\textsc{RR}}}
\newcommand{\ES}{\textit{\textsc{ES}}}
\newcommand{\LA}{\textit{\textsc{LA}}}
\newcommand{\AR}{\textit{\textsc{AR}}}
\newcommand{\MC}{\textit{\textsc{MC}}}

\newcommand{\roletrans}{\hookrightarrow_r}
\newcommand{\typetrans}[1]{\xhookrightarrow{#1}_t}

\newcommand{\ds}{\textsc{DepSearch}}
\newcommand{\DS}{Dependency Search}
\newcommand{\fds}{\textsc{\textit{f}}\ds{}}

\begin{document}

\title{Towards a Uniform Framework for Dynamic Analysis of Access Control Models}

\author{Peter Amthor}
\institute{Ilmenau University of Technology, Germany\\
\email{peter.amthor@tu-ilmenau.de}}

\maketitle

\begin{abstract}
Security-critical system requirements are increasingly enforced through mandatory access control systems. These systems are controlled by security policies, highly sensitive system components, which emphasizes the paramount importance of formally verified security properties regarding policy correctness. For the class of safety-properties, addressing potential dynamic right proliferation, a number of known and tested formal analysis methods and tools already exist. Unfortunately, these methods need to be redesigned from scratch for each particular policy from a broad range of different application domains.
In this paper, we seek to mitigate this problem by proposing a uniform formal framework, tailorable to a safety analysis algorithm for a specific application domain. We present a practical workflow, guided by model-based knowledge, that is capable of producing a meaningful formal safety definition along with an algorithm to heuristically analyze that safety. Our method is demonstrated based on security policies for the SELinux operating system.
\end{abstract}

  \section{Introduction}
\label{sec:intro}

In a wide range of modern applications, software systems engineers are challenged by tightening security requirements based on legal and economical requirements. This challenge has been met by the approach of policy-controlled systems, whose design and implementation relies on formally specified security policies \cite{Watson13a}. These policies define rules that, reliably enforced by a system's implementation, can be proven to meet formalized security requirements. One of the most important families of security policies is described by access control (AC) models, which can be used to model and verify most security policies of today's policy-controlled systems  \cite{Spencer99a, 
Loscocco01a, 
Watson03d, 
Smalley13a, 
Russello12a, 
Bugiel13a, 
Faden07a, 
Grimes07a}. 

Because of the paramount importance of formal models, a plethora of methods has emerged for their specification, analysis and implementation. More specifically, models for access control policies can be subdivided based on two principal objectives: either, to precisely specify the semantics of a particular policy family, e.\,g. roles \cite{Sandhu00a} or user relationships \cite{Fong11b, Crampton14a}, or to focus on a precise definition of a particular security property, which represents a formalized security requirement (such as confinement of access right proliferation \cite{Harrison76a, Ferrara13b} or information flows \cite{Kafura13a}). However, models resulting from both objectives have turned out incompatible due to different formalisms; moreover, any new model focused on either objective has always required a new formal calculus from scratch.

These problems have been addressed by the paradigm of model-based security engineering \cite{Barker09a, Poelck11a, Kafura13a, Amthor14a, Poelck14a}. Its goal is a uniform pattern for designing and analyzing security models, which should encompass the full bandwidth of practically important analysis goals, including dynamic, time-variant security properties in the lineage of HRU safety \cite{Harrison76a}. Aiming at these requirements, a recent, promising formal framework is the core-based modeling pattern \cite{Poelck11a, Poelck14a}.

To streamline model-based systems security engineering based on this pattern, we have introduced the entity labeling approach as a complementary pattern \cite{Amthor16a}. We have shown how access control policies of the SELinux operating system, as a typical representative of policy-controlled systems, can be modeled using standardized, uniform semantic abstraction based on the principle of attribution. The resulting SELinux access control model \SELX{} was discussed with a focus on specification costs, nevertheless neglecting an in-depth review of the actual policy analysis merits and costs.

In this work, we argue that both the core-based and the entity labeling pattern pave the way towards a uniform model analysis framework. This claim is substantiated by demonstrating how to tailor a general, heuristic-based analysis method for dynamic system properties (commonly known as \emph{safety} properties) to the \SELX{} model by leveraging the additional semantic information. In the process, we will define fundamental abstract interfaces to safety analysis methods that can be easily reused to tailor the same approach to other policy domains.

Contributions of this work include
\begin{enumerate}
	\item a generalization of the \ds{} heuristic for model safety analysis \cite{Amthor13a}, with the goal to be adaptable to every core-based/EL access control model,
	\item a partially automatable workflow of tailoring both a formal safety property and an appropriate heuristic to a specific model,
	\item a demonstration of this process for \SELX{}, resulting in a specialized safety analysis algorithm for SELinux security policies.
\end{enumerate}

The rest of this paper is organized as follows: After a brief discussion of relevant related work (Sec.~\ref{sec:relwork}), both the core-based and the entity labeling pattern for model-based security engineering are described (Sec.~\ref{sec:modelingpatterns}). This provides a basis for reviewing a promising heuristic approach to dynamic model analysis, \ds{}, in Sec.~\ref{sec:modelanalysis}, which is then rewritten in a generalized form (Sec.~\ref{sec:fds}). Sec.~\ref{sec:fds_tailoring} describes the novel workflow of tailoring this general heuristic to an actual access control model, which is finally applied to the practical example of SELinux in Sec.~\ref{sec:transformation}. We conclude with Section~\ref{sec:concl}.

  \section{Related Work}
\label{sec:relwork}

This work is based on our previous work towards flexible modeling patterns \cite{Amthor14a, Amthor16a}. A significant body of research has covered similar goals of unifying security models to reduce the costs of systems security engineering in practice. Notable work in this area includes meta models for abstract access control policies \cite{Barker09a, Barker10a}, information flow policies \cite{Kafura13a}, attribute-based access control policies \cite{Jin12a}, and diverse hybrid paradigms \cite{Ferraiolo11a, Huang12a}. All these approaches share the goal of unifying and thus streamlining policy specification and implementation; formal analysis of policies on the other hand is not regarded a focal goal in any of them. Consequently, a generalization of automaton-based model descriptions (\emph{core-based} modeling) was introduced to model vital system properties, such as PEP interfaces and protection state dynamics, in a uniform yet flexible calculus \cite{Poelck11a, Poelck14a}. Our approach to generalizing heuristics for safety analysis is based on this latter pattern.

Due to their paramount importance in practical attack scenario, security properties based on system dynamics have always received significant attention in terms of a broad body of formal analysis methods. Ultimately, however, the fundamental property of \emph{safety} (introduced by Harrison, Ruzzo and Ullman \cite{Harrison76a}) is proven to be undecidable -- which led to massive efforts in reducing a model's complexity and thus expressive power in a least-obstructive way, assuming a specific, well-defined application domain. In virtue of such expressive restrictions, dynamic model analysis approaches such as \cite{Tripunitara15a, Ranise14a, Sandhu88a, Sandhu92b, Ferrara13a, stoller11symbolic} share the goal of precisely deciding their particular descendant of \emph{safety}. Based on the goal of maximum flexibility, which naturally prohibits any restrictive assumptions about policy semantics, our approach takes the opposite path: instead of sacrificing parts of a model's expressive power, heuristic safety analysis sacrifices decidability. This enables tractability of the original, merely semi-decidable safety property and naturally yields an approximate solution.%
\footnote{An actually satisfied safety-requirement is not confirmable; however the critical case of unsatisfied safety, once confirmed, can be traced back to its cause.} 
Our most promising approach to heuristic analysis, \ds{} \cite{Amthor13a}, is therefore generalized in this paper to fit any core-based entity labeling model. The notion of safety used in our approach is based on Tripunitara's and Li's \cite{Tripunitara13a} precise and meaningful revisions of the original safety-definition.

Our approach tries to merge the merits of two general information modeling paradigms: attributes (\emph{ABAC} in the security context) and ontologies. The latter is leveraged in our approach to introduce semantic knowledge into the security policy engineering as early as possible, just before formalization starts, based only on a security requirements analysis. The goal is to reduce the potential of human error in the later phases of formal analysis and specification of a security policy, which should be automated based on rules that can be derived from the richer semantical knowledge about the model.

  \section{Model Engineering}
\label{sec:modelingpatterns}

This section introduces the two basic formal approaches we will use to model and analyze AC systems: the core-based modeling pattern by Pölck \cite{Poelck11a, Poelck14a}, and the entity labeling (EL) pattern \cite{Amthor15b, Amthor16a}. A security modeling pattern can be regarded a specialized type of ontology, defined along with a particular formal calculus. Each modeling pattern reflects a slightly different view on an AC system, tailored to different policy specification and/or policy analysis goals.

Throughout the rest of this paper, we will use the following \textbf{conventions for formal notation}: 
$\models$ is a binary relation between variable assignments and formulas in second-order logic, where $I \models \phi$ iff $I$ is an assignment of unbound variables to values that satisfies $\phi$. In an unambiguous context, we will write $\tup{x_0, \dots, x_n} \models \phi$ for any assignment of variables $x_i$ in $\phi$ that satisfies $\phi$. 
A logical formula assigned to a variable is delimited by $\logexpr{\nobreakspace}$. We will mark fixed values referenced from outside the formula by underlining, e.g. a fixed $x$ in an expression $\phi = \logexpr{y = \underline{x}}$.
For any mapping $f$, $f[x \mapsto y]$ denotes the mapping which maps $x$ to $y$ and any other argument $x^\prime$ to $f(x^\prime)$. 
For any mapping $f : A \rightarrow B$, $f\restriction_{A^\prime}$ denotes a restriction of $f$ to $A^\prime \subset A$ that maps any argument $x^\prime \in A^\prime$ to $f(x^\prime)$, whereas $f\restriction_{A^\prime}(x)$ is undefined for any $x \in A \setminus A^\prime$. 
For any set $A$, $2^A$ denotes the power set of $A$. 
$\mathbb{B}$ is the set of Boolean values $\top$ (\emph{true}) and $\bot$ (\emph{false}). 

\subsection{Core-based Modeling}
\label{sec:coremodeling}
\vspace*{-.1em}

The goal of the core-based model engineering paradigm is to establish a uniform formal basis for specification, analysis and implementation of diverse security models. 
A core-based access control model is defined as an extended state machine, called \emph{model core}:
\begin{equation*}
\tup{Q, \Sigma, \delta, \lambda, q_0, \Ext }
\label{eq:def_core}
\end{equation*}
where $Q$ is a (finite or infinite) set of protection states, $\Sigma$ is a (finite or infinite) set of inputs, $\delta : Q \times \Sigma \rightarrow Q$ is the state transition function, $\lambda : Q \times \Sigma \rightarrow \mathbb{B}$ is the output function, $q_0 \in Q$ is the initial protection state, and \Ext{} is an arbitrary tuple of static model extensions.
The model core can be tailored to any domain-specific security policy in terms of dynamic state members and static model extensions. We formally do this by introducing two sets $\mathit{DYN}$ of model components that may change their state, and $\mathit{STAT}$ of such which may not. In practice, these sets are determined by the semantics of security policy rules: For a classical AC policy equivalent to an HRU model for example, we will define $\mathit{DYN} = \{S, O, \mathit{acm}\}$ and $\mathit{STAT} = \{R\}$ for the model components subjects set $S$, objects set $O$, access control matrix $\mathit{acm} : S \times O \rightarrow 2^R$, and rights set $R$.

Based on the abstract automaton and the definition of model components, three steps are required to describe a particular AC system through a core-based model (cf. \citet[pp.~25 et seq.]{Poelck14a}): (1.) Specializing $Q$, i.e. explicitly defining the automaton's state space members (based on $\mathit{DYN}$). (2.) Specializing \Ext{}, i.e. defining static model components (based on $\mathit{STAT}$). (3.) Specializing $\delta$ and $\lambda$, i.e. describing the dynamic behavior of the AC system. Depending on step 1, the initial protection state $q_0$ has to be specified according to the particular analysis goal. Depending on both steps 1 and 2, the input alphabet $\Sigma$ has to be specified according to the interface of the modeled access control system.

In step 3, protection state dynamics are described by the state transition function $\delta$ through pre- and post-conditions of every possible state transition. This is done by comparing each input with two formulas in second-order logic, \PRE{} and \POST{}. We then define $\delta$ by formally specifying the conditions that each pair of states $q$ and $q^\prime$ has to satisfy w.r.t. an input $\sigma \in \Sigma$ for a state transition from $q$ to $q^\prime$ to occur:
\begin{equation*}
\delta(q, \sigma) = \left\{	\begin{array}{cl}
							q^\prime, & 	\tuple{q, \sigma} \models \PRE{}	\wedge\, \tuple{q^\prime, \sigma} \models \POST{}\\
							q,        & \mbox{otherwise.}
							\end{array}\right.
\label{eq:def_delta}
\end{equation*}
Because an access control system is fundamentally deterministic, \POST{} requires that $q^\prime$ equals $q$ where not redefined. In practice, \PRE{} and \POST{} are divided into \emph{commands} that match particular interface calls of the modeled system: For any command $\mathit{cmd} \in \Sigma_C$, $\PRE{}(\mathit{cmd})$ denotes the partial, command-specific pre-condition of $\mathit{cmd}$ and $\POST{}(\mathit{cmd})$ its post-condition. The AC system's interface is then modeled by $\Sigma = \Sigma_C \times \Sigma_X$, where $\Sigma_C$ is a set of command identifiers and $\Sigma_X$ contains sequences of possible values (command parameters) for variables in \PRE{} and \POST{}. This engineering-friendly notation dates back to the first access control models \cite{Bell76a, Harrison76a} and allows for a structured, system-specific notation of $\delta$ in terms of partial definitions depending on each input command. We will call the set of such partial definitions $\Delta = \{ \tup{\mathit{cmd}, x_\mathit{cmd}, \PRE{}(\mathit{cmd}), \POST{}(\mathit{cmd})} | \mathit{cmd} \in \Sigma_C; x_\mathit{cmd} \in \Sigma_X; \PRE{}(\mathit{cmd}), \POST{}(\mathit{cmd}) \in \mathbb{B} \}$ a model's \emph{state transition scheme}, a specification of the behavior of $\delta$.
The notation of a simple example command \emph{delegateRead} of an HRU model's state transition scheme is shown in Fig.~\ref{fig:example_cmd}.

Finally, to describe authorization decisions of an AC system, the automaton features an output function $\lambda$. It enables the analysis of correct policy behavior and thus supports a formally verified specification. $\lambda$ defines a binary access decision $\lambda(q, \sigma) \Leftrightarrow \tuple{q, \sigma} \models \PRE{}$.

\vspace*{-.5em}
\begin{figure*}[thbp]
\centering
\begin{minipage}[t]{0.9\textwidth}
\begin{tabular}{rl}
\multicolumn{2}{l}{$\blacktriangleright$ \textbf{delegateRead}($s_1, s_2, o$) $::=$}\\
\PRE:			& $\mathit{read\_right} \in \mathit{acm}_q(s_1, o)$\,;\\
\POST:			& $\mathit{acm}_{q^\prime} = \mathit{acm}_q[\tup{s_2, o} \mapsto \mathit{acm}_q(s_2, o) \cup \{\mathit{read\_right}\}]$
\end{tabular}\\[.75ex]
\end{minipage}
\caption{Exemplary command definition for a core-based HRU model. We specify a command $\mathit{delegateRead}$ with parameters $\tup{s_1, s_2, o}$ that models delegation of $\mathit{read\_right}$ regarding $o$ by $s_1$ to $s_2$. $\PRE{}(\mathit{delegateRead})$ expresses that $s_1$ needs to possess this right for a state transition by this command to occur, $\POST{}(\mathit{delegateRead})$ expresses that $s_2$ needs to do so afterwards.}
\label{fig:example_cmd}
\vspace*{-1.5em}
\end{figure*}

\subsection{Entity Labeling}
\label{sec:elmodels}
\vspace*{-.1em}

Entity Labeling (EL) is an abstract semantic modeling pattern for the formalization of contemporary access control policies. It was originally introduced with a focus on operating systems security policies \cite{Amthor15b, Amthor16a} and is based on three observations regarding the semantics of such policies:

\begin{enumerate}
	\item attributes that label entities are used for making access decisions
	\item the protection state described is dynamic, i.\,e. the actual system's configuration that policy rules refer to may change over time
	\item time-invariant constraints restrict possible changes of the protection state; in practice, such constraints may be system-intrinsic (e.\,g. policy consistency criteria) as well as external (responding so a system's environmental context)
\end{enumerate}

These observations lead to six semantic categories that describe an access control system from a quite different angle than core-based modeling does: as an ontology describing attribution-based policy logic. In practice, core-model components in $\mathit{DYN}$ and $\mathit{STAT}$ also fall into one of these EL categories:

\begin{description}
	\item[Label Set (\LS):] Set of legal label values.
	\item[Relabeling Rule (\RR):] Rules for legal label changes.
	\item[Entity Set (\ES):] Sets of entity identifiers.
	\item[Label Assignment (\LA):] Associations between a entity and its labels.
	\item[Access Rule (\AR):] Rules that describe, given two or more labels, operations that accordingly labeled entities are allowed to perform. AR components typically constitute an AC model's access control function (ACF).
	\item[Model Constraints (\MC):] Constraints over the other components that must be satisfied in every model state.
\end{description}
For specializing these categories, their semantics have to be matched to policy logic -- a task that should be performed based on a system's security requirements, ideally during requirements analysis and system design. This highlights the importance of a rigorous requirements engineering, which lays the foundation of later model-based security analysis. At the same time, EL modeling helps to introduce and preserve such critical knowledge about an AC system's intended semantics early in the whole engineering process.

On a formal level, the EL pattern is described by a tuple $\tup{\mathit{CAT}, \mathit{sem}}$
where $\mathit{CAT} = \{\mathit{LS}, \mathit{RR}, \mathit{ES}, \mathit{LA}, \mathit{AR}, \mathit{MC}\}$ is the set of EL category identifiers and $\mathit{sem}: \mathit{CAT} \rightarrow 2^{\mathit{DYN} \cup \mathit{STAT}}$ is their semantics association to model components as described above.
For the sake of brevity, EL semantics of a model \textsf{M} are denoted by
$
\mathsf{M}|_\mathit{cat} = \mathit{sem}(\mathit{cat})
$
where $\mathit{cat} \in \mathit{CAT}$ is a semantic category identifier. 

Note that the EL pattern complements core-based modeling by adding abstracted knowledge about a system's requirements to the automaton-based calculus. Using both in combination, we can reason about meaningful analysis goals, their formal definition, and the interpretation of analysis results (e.\,g. the semantic origins of a right proliferation in a policy's logic) as part of a generalizable security engineering approach. We argue for this claim by presenting a generalized analysis approach for dynamic model properties in the following sections.

  \section{Model Analysis}
\label{sec:modelanalysis}

In this section, we discuss the basic idea of heuristic safety analysis. We will outline one of the most successful algorithmic approaches to this problem, \ds{}, presented for the HRU access control model in \citet{Amthor13a, Amthor14a}. On this basis, we will later introduce a systematic generalization of \ds{}.

\subsection{Heuristic Safety Analysis}
\label{sec:heuristic}

The most general research questions targeted by the analysis of an access control system relate to the proliferation of access rights. As formalized in the seminal HRU access control model \cite{Harrison76a}, the \emph{safety} of a system fundamentally addresses such questions: Given a protection state of an HRU model, is it possible that some subject ever obtains a specific right with respect to some object? If this may happen, such a model state is considered \emph{unsafe} with respect to that right. An intuitive yet practically meaningful interpretation of this question, \emph{(r)-simple-safety}, has been defined (via its complement) by Tripunitara and Li \cite{Tripunitara13a}:
\begin{definition}
Given a core-based HRU model $\tup{Q, \Sigma, \delta, \lambda, q_0, \tup{R} }$, a state $q = \tup{S_q, O_q, \mathit{acm}_q} \in Q$ is \mbox{\bf (r)-simple-unsafe} with respect to a right $r \in R$ iff $\exists\, q^\prime = \tup{S_{q^\prime}, O_{q^\prime}, \mathit{acm}_{q^\prime}} \in \{ \delta^\ast(q, a) \,|\, a \in \Sigma^\ast \}$:
\begin{equation*}
\begin{array}{rl}
& \exists s \in S_{q^\prime}, \exists o \in O_{q^\prime} : r \in \mathit{acm}_{q^\prime}(s, o)\\
\wedge & \big( s \notin S_q \vee o \notin O_q \vee r \notin \mathit{acm}_q(s, o) \big) \, .
\end{array}
\end{equation*}
\vspace*{-1em}
\label{def:r-simple-unsafe}
\end{definition}
where $\delta^\ast : Q \times \Sigma^\ast \rightarrow Q$ is the transitive state transition function defined as $\delta^\ast(q, \sigma \circ b) = \delta^\ast(\delta(q,\sigma),b)$ and $\delta^\ast(q, \epsilon) = q$ for any $\sigma \in \Sigma \cup \{ \epsilon \}, b \in \Sigma^\ast$.

Two facts are worth noting here: First, \emph{safety} as per Def.~\ref{def:r-simple-unsafe} always relates to both a specific model state $q$ to analyze (in practice, this is a momentary configuration of the system in question) and a specific access right $r$ whose proliferation we are interested in. We call this $r$, being the source of an HRU model's authentication mechanics, a safety analysis \emph{target}. Second, Def.~\ref{def:r-simple-unsafe} is least restrictive with respect to $r$ since it allows \emph{any} subject or object to violate safety. In practice, possible analysis questions may rather concentrate on a specific $s$ or $o$, leading to more specialized safety definitions such as \emph{(s,o,r)-simple-safety} \cite{Tripunitara13a}.

It is long since known that any variant of safety is not decidable, given an access control models with unrestricted expressive power and thus infinite state space \cite{Harrison76a}. 
In order to analyze such models, we must trade accuracy for tractability: using a heuristic algorithm, unsafe model states may be found (given such exist), while termination of the algorithm cannot be guaranteed. The idea of heuristic safety analysis thus leverages the semi-decidability of the problem.
On the plus side, valuable hints on model correctness are obtained if unsafe states are found and policy engineers are pointed to input sequences that lead to such states. 

The strategy behind heuristic safety analysis algorithms is to find an input sequence that, starting at $q$, enters $r$ into a matrix cell of some follow-up state $q_{target}$. When this happens, $q$ is proven to be unsafe with respect to $r$; as long as no such target state is found, the search continues. Therefore, a successful algorithm must exploit model properties that maximize the probability of an input to contribute to a path from $q$ to $q_{target}$.

\subsection{\DS{}}
\label{sec:depsearch}

We will now outline the \ds{} safety analysis heuristic for HRU models. We developed it based on the insight that in the most difficult case, right leakages in a model are well hidden and appear only after long state transition sequences where each command executed depends exactly on the execution of its predecessor. Even for such hard analysis cases, \ds{} has turned out successful \cite{Amthor13a}. Essentially, \ds{} consists of two phases: static and dynamic analysis. We will discuss these phases on an informal basis, for an in-depth discussion and evaluation of the algorithm see \citet{Amthor13a, Amthor14a}.

In the first phase, a \textbf{static analysis} of the HRU state transition scheme is performed. It yields a structured description of inter-command dependencies, constituted by entering (as a part of \POST{}) and requiring (part of \PRE{}) the same right in two different commands. The knowledge about these dependencies is encoded in a \emph{command dependency graph} (CDG) whose vertices are commands, and an edge from command $c_1$ to command $c_2$ denotes that a post-condition of $c_1$ matches at least one pre-condition of $c_2$.

The CDG is assembled in a way that all paths from vertices without incoming edges to vertices without outgoing edges indicate input sequences for reaching $q_{target}$ from $q$. To achieve this, two virtual commands $c_q$ and $c_\mathit{target}$ are generated: $c_q$ is the source of all paths in the CDG, since it represents the state $q$ to analyze in terms of a command specification added to $\Delta$. It is generated by encoding $\mathit{acm}_q$ in $\POST{}(c_q)$. In a similar manner, $c_\mathit{target}$ is the sink of all paths in the CDG, which represents all possible states $q_{target}$ by checking the presence of the target right in any matrix cell in $\PRE{}(c_\mathit{target})$.

In the second, \textbf{dynamic analysis} phase, the CDG is used to guide dynamic state transitions by generating input sequences to the automaton. The commands involved in each sequence are chosen according to different paths from $c_q$ and $c_\mathit{target}$. Consequently, \ds{} successively generates input sequences by traversing the CDG on every possible path and in turn parameterizing the emerging sequence of commands with values that can be inferred from a constraint satisfaction problem (CSP) solver. Each effected state transition is simulated by the algorithm, and once a CDG path is completed, the validity of the unsafety-criteria (Def.~\ref{def:r-simple-unsafe}) is checked.

  \section{Generalized Framework}
\label{sec:fds_framework}

As introduced so far, the \ds{} algorithm is restricted to analyzing models of the classical HRU calculus. The family of dynamic model properties however spans the whole range of access control models, which requires an adaption of \ds{} for each an every such model. 
In \citet{Amthor14a}, we have hinted at a uniform generalization of \ds{}, however, the previous work always resorts to HRU; problems related to tailoring a general core-based model to a specific notion of safety and interpreting this in terms of \ds{} have not yet been addressed in detail. In particular, there was no uniform workflow for heuristic-tailoring that avoids a re-design of each mode-dependent part of the heuristic.

With entity labeling we have introduced a plus of semantic information that can be used to tailor such a general safety analysis algorithm to a specific model and analysis goal in a consistent workflow. In this section, we will first present the idea behind the generalization, fallowed by a description of the workflow how to specialize it. This workflow will then be demonstrated based on the practical case of an SELinux access control model in Sec.~\ref{sec:transformation}.

\begin{algorithm}[tbp]
\KwIn{
	$\delta$ \dots{} model's state transition function \newline
	$\Delta$ \dots{} model's state transition scheme, specifying $\delta$ \newline
	$q_0$ \dots{} model state the safety of which is to be analyzed \newline
	$\mathit{target}$ \dots{} leakage target
}
\KwOut{
	$\mathit{STS}$ \dots{} state transition sequence leaking $\mathit{target}$
}
\BlankLine
$q \leftarrow q_0$\;
$\tup{\mathit{CDG}, c_q} \leftarrow$ CDGAssembly($\Delta,q,\mathit{target}$)\;
$\mathit{STS} \leftarrow q$\;
\BlankLine
\Repeat{\nlset{(a)} 
   $\mathrm{isLeaked}(q_0, q^\prime, \mathit{target})$}{
  $\mathit{path} \leftarrow$ CDGPathGeneration($\mathit{CDG}, c_q$)\;
  \nlset{(d)}
  $\mathit{params} \leftarrow \mathrm{assignParams}(q, \mathit{path})$\;
  \While{$c \leftarrow \mathit{path.nextNode}$}{
    $q^\prime \leftarrow \delta(q, c, \mathit{params(c)})$\;
    $\mathit{STS} \leftarrow \mathit{STS} \circ q^\prime$\;
    $q \leftarrow q^\prime$\;
    }
}
\Return{$\mathit{STS}$}\;
\caption{\fds{}}
\label{alg:fds}
\end{algorithm}

\begin{algorithm}[tbp]
\KwIn{
	$\Delta$ \dots{} model's state transition scheme \newline
	$q$ \dots{} model's state the safety of which is analyzed \newline
	$\mathit{target}$ \dots{} leakage target
}
\KwOut{
	$\tup{V, E}$ \dots{} command dependency graph \newline
	$c_q$ \dots{} starting point for the command sequence generation
}
\BlankLine
\Proc({predecessors(\textbf{in} $v \in V$)}){
  	\nlset{(c)}
	$P \leftarrow \mathrm{buildPredSet}(\Delta, v)$\;
	\For{$c \in P$}{
		\If{$c \notin V$}{
		$V \leftarrow V \cup \{c\}$\;
		predecessors($c$)\;
		}
	$E \leftarrow E \cup \{\tup{c,v}\}$\;
	}
}
\BlankLine
\nlset{(e)}
$c_q \leftarrow$ createCDGSource($q$)\;
\nlset{(b)}
$c_{target} \leftarrow$ createCDGSink($\mathit{target}$)\;
$\Delta \leftarrow \Delta \cup \{c_q\}$\;
$V \leftarrow \{c_{target}\}$\;
$E \leftarrow \emptyset$\;
predecessors$(c_{target})$\;
\Return{$\tup{V, E}, d, \hat{d}, c_q$}\;

\caption{\fds{}::CDGAssembly}
\label{alg:fdsCdgAssembly}
\end{algorithm}

\subsection{Generalizing \ds{}}
\label{sec:fds}

\begin{figure}
\centering
\begin{tikzpicture}[grow'=right]
\tikzset{level distance=97pt,sibling distance=0pt}
\tikzset{execute at begin node=\strut}
\tikzset{every tree node/.style={align=left}}
\Tree [.\fds{} 
		[.{(Model-Independent\\ Modules)}
			CDGPathGeneration
		]
		[.{(Model-Specific\\ Modules)}
			[.CDGAssembly 
				{\itshape buildPredSet} {\itshape createCDGSource} {\itshape createCDGSink}
			]
			{\itshape assignParams}
			{\itshape isLeaked}
		]
	]
\end{tikzpicture}
\caption{Modules in \fds{}. Generic interfaces are printed italic.}
\label{fig:fds_modules}
\end{figure}
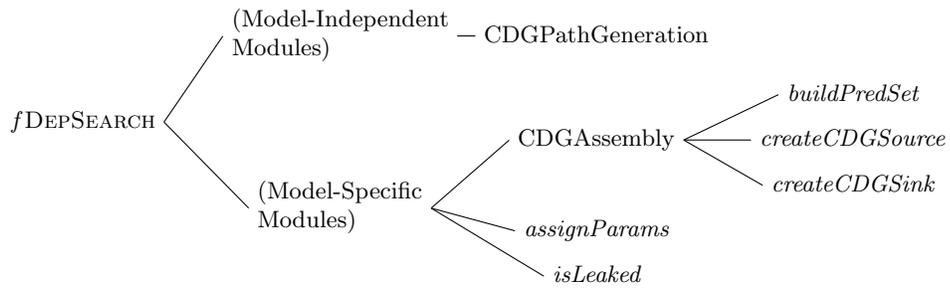

In this section, we present \fds{} as a variant of the original HRU algorithm including a number of abstract interfaces. These interfaces originate from a modular view on the heuristic, which is composed of both model-independent and mode-dependent modules. Fig.~\ref{fig:fds_modules} illustrates this. 
Based on the model-dependent submodules, we consider three semantic abstractions: \emph{safety}, \emph{dependency}, and \emph{model dynamics}. These abstractions and their related interfaces will now be discussed based on the \fds{} algorithm specifications (Algs.~\ref{alg:fds} and \ref{alg:fdsCdgAssembly}).\footnote{We have stripped down the algorithm specification to the relevant parts for this discussion, leaving aside path generation strategies and graph traversal attributes, which are exclusively model-independent and thus need not to be tailored.} 

The interfaces \emph{isLeaked} (Alg.~\ref{alg:fds}\,(a)) and \emph{createCDGSink} (Alg.~\ref{alg:fdsCdgAssembly}\,(b)) represent the analysis goal of the heuristics, expressed through a formal definition of \textbf{safety} (such as Def.~\ref{def:r-simple-unsafe} for HRU) and its target in terms of some model component (such as a right $r$). \emph{isLeaked} is specified as a boolean function that implements a formal unsafety criteria, which may be satisfied by any model state $q^\prime$ reached during heuristic execution. \emph{createCDGSink} is part of the static model analysis and uses the knowledge about the safety target to generate a virtual command $c_\mathit{target}$ in the CDG as described in Sec.~\ref{sec:depsearch}, which ensures that generated paths during the dynamic analysis always end in a command leaking the target.

Implementing the safety abstractions also enables an implementation of the model's notion of \textbf{dependency} through the interfaces \emph{buildPredSet} (Alg.~\ref{alg:fdsCdgAssembly}\,(c)) and \emph{assignParams} (Alg.~\ref{alg:fds}\,(d)). The former steers the generation of the CDG during static model analysis, where a formal definition of dependency (based on the previously defined safety and target) is used to add predecessor-edges between a command and other commands that establish necessary conditions for it to be executed. In an HRU model for example, this means for any given command $\mathit{cmd}$ in the state transition scheme, a command $\mathit{cmd}^\prime$ is a predecessor of $\mathit{cmd}$ iff entering rights in $\POST{}(\mathit{cmd}^\prime)$ has an impact on the value of $\PRE{}(\mathit{cmd})$.

The latter interface, \emph{assignParams}, is actually another heuristic in itself. Since the static phase of \ds{} analyzes dependencies based on \PRE{} and \POST{} formulas only, yet disregarding the actual assignment of parameters that determine their values, parameter selection becomes a problem on its own. Again, the goal here is to maximize satisfiability and thus execution of commands on a CDG path that may eventually leak the safety target. Therefore, an implementation of \emph{assignParams} must also take into account dependencies between commands, which may now established by their parameters relating to common model components. We have developed an approach to express the parameter selection problem as a constraint satisfaction problem by introducing a second graph, the parameter constraint network, that is input to a heuristically modified version of Poole's and Mackworth's arc consistency algorithm \cite{Mackworth10a}. An evaluation of this parameter selection heuristic is still ongoing work. Formally, \emph{assignParams} is an interface that evaluates both a base state $q$ for path execution and the definition of commands in the path, to return a mapping of these commands to assignments for their individual parameters.

As a last abstract interface, \emph{createCDGSource} (Alg.~\ref{alg:fdsCdgAssembly}\,(e)) generates the virtual command $c_q$ in the CDG. Its implementation is independent from the other abstractions, since it only takes into account knowledge of \textbf{model dynamics} to construct $\POST{}(c_q)$ based on the model state the safety of which is to be analyzed.

\subsection{Tailoring \fds{}}
\label{sec:fds_tailoring}

In practice, manual implementation of these interfaces from scratch is still an error-prone process. To support this step and thus streamline the model-independent use of \fds{}, EL categories are used to guide a three-phases workflow: First, the actual core-based model is created and amended by EL semantics. Second, a meaningful formal safety definition is derived from the model. Third, the abstract interfaces are actually implemented based on the safety definition and additional information from EL semantics. We will now detail these phases and guidelines they leverage for a well-structured, error-reducing manual or even partially automated process of dynamic model analysis.

\subsubsection{1. Model Specification}

Goal of the first phase is to model an access control policy in a formal calculus suitable for dynamic analysis.

\textbf{(Step 1.1)} 
As soon as an AC policy's semantics are designed, EL categories should be used to identify and classify formal structures such as sets, relations, and mapping according to Sec.~\ref{sec:elmodels}. Result is an EL model $\tup{\mathit{CAT}, \mathit{sem}}$. 

\textbf{(Step 1.2)} 
In \cite[pp.\,25--27]{Poelck14a}, the basic steps of specializing a core-based model for an access control policy are outlined (we summarized them in Sec.~\ref{sec:coremodeling}). In order to enable the modeling of dynamic behavior, these steps should now be applied to our EL model. We therefore classify model components into $\mathit{DYN}$ and $\mathit{STAT}$ and derive a state transition scheme $\Delta$ from policy rules.\footnote{A detailed example an how the latter can be obtained in practice is given in \citet{Amthor16a}.} Result is a second, core-based model $\mathsf{M} = \tup{Q, \Sigma, \delta, \lambda, q_0, \Ext }$ covering the same policy. 

\subsubsection{2. Safety Specification}

In the second phase, the analysis goal has to be formalized in terms of a safety-property.

\textbf{(Step 2.1)} 
The access control function describes the impact of model components on the dynamic behavior of $\lambda$. Therefore, it formally defines the necessary conditions for any kind of safety-violating state transition (i.\,e. leakage). In practice, the ACF of a system may by obtained from two different sources: On the one hand, requirements engineering on a rigorously low level of abstraction may have already stipulated access restrictions in a formal way. On the other hand, especially in case of rich and complex policies that steer diverse security mechanisms,\footnote{In today's operating systems for example, a hybrid of attribute-based access control and information flow policies is common.} a condensed and precise ACF may not be achievable right from the beginning of system engineering. In this latter case however, EL model semantics can be used to bridge the gap between informal requirements and a formal ACF definition as follows.

Formally, an ACF is a function $\mathit{acf} : \Sigma \rightarrow \mathbb{B}$ with $\mathit{acf}(\sigma) \Leftrightarrow \phi_\mathit{acf}$. 
Since the occurrence of model components from \emph{AR} in $\PRE{}(c)$ is a necessary condition for a command $c$ to make an authorization decision (cf. EL semantics), a disjunction over such conditions determines the ACF as follows:
\begin{equation}
\begin{array}{rcl}
\phi_\mathit{acf} 	&::=& \logexpr{\bigvee_{\phi_i \in \underline{\Phi_\mathit{AR}}} \phi_i} \mbox{ where }\\
\Phi_\mathit{AR}  	&::=& \{ \PRE{}(c), \mathit{c} \in \Sigma_C \,|\, \exists x \in \mathsf{M}|_\mathrm{AR} : \\
					& 	& \PRE{}(c) \mbox{ depends on the value of } x \} \, .
\end{array}
\label{eq:def_acf-construction}
\end{equation}

\textbf{(Step 2.2)} 
Based on the ACF, we will now extract such clauses involving model components in the set $\mathit{DYN}$ from the core-based pattern: this yields a set of dynamically changing leakage conditions. 
Each possible effect of a state transition influencing one of these clauses can be considered a leak, which leads to a maximal number of safety definitions based on these leaks. These definitions together with the target model components can be inferred automatically, as demonstrated in Sec.~\ref{sec:eval-safety}.

\textbf{(Step 2.3)} 
For a complex model featuring multiple components responsible for authorization, it becomes obvious that a number of alternate safety definitions is formally possible. Which one of them however is meaningful in practice must be decided in the next step and requires human knowledge, possibly once again by comparison against security requirements. Model components from the EL category RR are responsible for authorization changes in a protection state and therefore give a first indicator for filtering irrelevant safety definitions. More details on a practical case will be presented in Sec.~\ref{sec:eval-safety}.

\subsubsection{3. Heuristic Specification}

In the last phase, the abstract interfaces of \fds{} can be implemented based on the formal model and the definitions of safety and analysis target. As outlined in Sec.~\ref{sec:fds}, the safety and target definitions guide the implementations of \emph{isLeaked} and \emph{creadeCDGSink} directly and, indirectly via the indirection of dependency abstraction drawn from the safety property, also the implementation of \emph{buildPredSet} and \emph{assignParams}. The most straight-forward implementation is that of \emph{createCDGSource}, based on the model dynamics declared through $\mathit{DYN}$. Again, Sec.~\ref{sec:eval-safety} will illustrate this process by implementing some of these interfaces.
  \section{Application to SELinux}
\label{sec:transformation}

The goal of this section is to demonstrate the tailoring of \fds{} based on a real-world case study. To this end, we specify a safety analysis algorithm for the widespread policy-controlled operating system SELinux. 
For the sake of clarity, we will restrict our study to the analysis goal of safety analysis and cover the EL modeling of our policy only to that extend; more specifically, we will omit the model constraints (\emph{MC}) components which are intended to support the verification of static model consistency properties. As used in SELinux, these model components could be used for proving interative policy properties similar to BLP security \cite{Bell76a}.

In the following sections, we will successively address the three major phases in tailoring our analysis framework: model specification, safety specification, and heuristic specification as described in Sec.~\ref{sec:fds_tailoring}.

\subsection{SELinux Model}
\label{sec:eval-model}
\vspace*{-.1em}

To set the stage for safety analysis, we will briefly revisit the core-based/EL model for the SELinux access control system (\SELX{}) from \citet{Amthor16a}. The model components that formally represent the SELinux access control semantics are
\begin{itemize}
	\item $C$: the set of object classes
	\item $U$: the set of SELinux users
	\item $R$: the set of roles
	\item $T$: the set of types
	\item $E$: the set of all system resources, including processes
	\item $\roletrans\: \subseteq R^2$: a relation modeling allowed role transitions
	\item $\typetrans{}\: \subseteq T^3$: a relation modeling entrypoints and allowed type transitions
	\item $P$: the set of permissions
	\item $\mathit{allow} : T \times T \times C \rightarrow 2^P$: a mapping of keys for TE-allow-rules to their associated permissions
	\item $\mathit{cl} : E \rightarrow C$: a mapping of resources to their object classes
	\item $\mathit{con} : E \rightarrow U \times R \times T$: a mapping of resources to their security contexts
\end{itemize}
These model components are classified according to both patterns as shown in Tab.~\ref{tab:el-selx}.

\begin{table}[tbp]
\centering
\caption{Classification of \SELX{} model components in EL and core-based modeling patterns.}
\begin{tabular}{ccc}
	\hline
	\noalign{\smallskip}
			\hspace*{.5em} \textbf{\SELX{}}   \hspace*{.25em}	
		&	\hspace*{.25em} $Q$ Members    \hspace*{.25em}	
		&	\hspace*{.25em} \Ext{} Members \hspace*{.5em}	\\
	\noalign{\smallskip}
	\hline 
	\noalign{\smallskip}
	\LS	&	---			&	$C, U, R, T$	\\
	\RR	&	---						&	$\roletrans, \typetrans{}$\\
	\ES	&	$E$				&	---\\
	\LA	&	$\mathit{cl}, \mathit{con}$	&	---\\
	\AR	&	---						&	$\mathit{allow}, P$\\
	\hline
\end{tabular}
\label{tab:el-selx}
\end{table}

\subsection{SELinux Safety}
\label{sec:eval-safety}

In \cite{Amthor16a}, it was argued that a two-step approach to specifying dynamic model behavior ($\delta$ and $\lambda$ of a core-based model) reduces analysis complexity and streamlines a generic approach to dynamic analyses. We will now substantiate this claim by using \emph{basic commands} for deriving a safety definition that allows to tailor \fds{} to this model. 
The definitions of \SELX{} basic commands to create and remove system resources, relabel processes and authorize access requests are given in Fig.~\ref{fig:selx_bcmd-defs}.

\begin{figure*}[thb]
\centering
\begin{minipage}[t]{0.4\textwidth}
\vspace{0pt}
\begin{tabular}{rl}
\multicolumn{2}{l}{$\blacktriangleright$ \textbf{create}($e, e^\prime, c^\prime$) $::=$}\\
\PRE:			& $e \in E_q$\\
$\wedge$			& $e^\prime \in E \setminus E_q$\\
$\wedge$			& $c^\prime \in C$\\
$\wedge$			& $\mathit{con}_q(e) = \tup{u, r, t}$\,;\\
\POST:			& $E_{q^\prime} = E_q \cup \{ e^\prime \}$\\
$\wedge$			& $\mathit{cl}_{q^\prime} = \mathit{cl}_q[e^\prime \mapsto c^\prime]$\\
$\wedge$			& $\mathit{con}_{q^\prime} = \mathit{con}_q[e^\prime \mapsto \tup{u, r, t}]$\\
\end{tabular}\\[.75ex]
\begin{tabular}{rl}
\multicolumn{2}{l}{$\blacktriangleright$ \textbf{remove}($e$) $::=$}\\
\PRE	:			& $e \in E_q$\,;\\
\POST:			& $E_{q^\prime} = E_q \setminus \{ e \}$\\
$\wedge$			& $\mathit{cl}_{q^\prime} = \mathit{cl}_q\restriction_{E_{q^\prime}}$\\
$\wedge$			& $\mathit{con}_{q^\prime} = \mathit{con}_q\restriction_{E_{q^\prime}}$\\
\end{tabular}%
\end{minipage}%
\hspace{10pt}%
\begin{minipage}[t]{0.4\textwidth}
\vspace{0pt}
\begin{tabular}{rl}
\multicolumn{2}{l}{$\blacktriangleright$ \textbf{relabel}($e, f, r^\prime, t^\prime$) $::=$}\\
\PRE:			& $e \in E_q$\\
$\wedge$			& $\mathit{cl}_q(e) = \mbox{process}$\\
$\wedge$			& $\mathit{con}_q(e) = \tup{u, r, t}$\\
$\wedge$			& $\mathit{con}_q(f) = \tup{\wc{}, \wc{}, \mathit{tf}}$\\
$\wedge$			& $r \roletrans r^\prime$\\
$\wedge$			& $t \typetrans{\mathit{tf}} t^\prime$\,;\\
\POST:			& $\mathit{con}_{q^\prime} = \mathit{con}_q[e \mapsto \tup{u, r^\prime, t^\prime}]$\\
\end{tabular}\\[.75ex]
\begin{tabular}{rl}
\multicolumn{2}{l}{$\blacktriangleright$ \textbf{access}($e, e^\prime, p$) $::=$}\\
\PRE:			& $\{e, e^\prime\} \subseteq E_q$\\
$\wedge$			& $\mathit{cl}_q(e) = \mbox{process}$\\
$\wedge$			& $\mathit{cl}_q(e^\prime) = c^\prime$\\
$\wedge$			& $\mathit{con}_q(e) = \tup{\wc{}, \wc{}, t}$\\
$\wedge$			& $\mathit{con}_q(e^\prime) = \tup{\wc{}, \wc{}, t^\prime}$\\
$\wedge$			& $p \in \mathit{allow}(t, t^\prime, c^\prime)$\,;\\
\POST:			& $\top$\\
\end{tabular}\\[.75ex]
\end{minipage}
\caption{\SELX{} basic commands of a low-level SELinux state transition scheme. For better readability we have replaced solely syntactical declarations of variables by the wildcard symbol ``\wc{}'' (which should otherwise be a generic, $\exists$-quantified placeholder).}
\label{fig:selx_bcmd-defs}
\vspace*{-1em}
\end{figure*}

\textbf{(Step 2.1)} 
To determine the \SELX{} ACF, only conditions checking \emph{AR} model components need to be taken into account. For example, $\PRE{}(\mathit{remove}) = \logexpr{e \in E_q}$ in Fig.~\ref{fig:selx_bcmd-defs} checks for the presence of a given entity $e$, which is independent of \emph{AR} model components and therefore constitutes a mere consistency-check rather than an authorization decision. 
In case of our \SELX{} basic commands, only \emph{access} is influenced by \emph{AR} components (which hints at the significance of a multi-level commands specification). Since $\mathit{acf}$ solely depends on $\PRE{}(\mathit{access})$, $\phi_\mathit{acf}$ from Equation~\ref{eq:def_acf-construction} expands to
\begin{equation}
\begin{array}{rcl}
\mathit{acf}(e, e^\prime, p)	& \Leftrightarrow	& \exists \, t, t^\prime, c^\prime : \\
							& \mbox{\scriptsize (\theequation.1)}	& \{e, e^\prime\} \subseteq E_q \\
							& \mbox{\scriptsize (\theequation.2)}	& \wedge \; \mathit{cl}_q(e) = \mbox{process} \\
							& \mbox{\scriptsize (\theequation.3)}	& \wedge \; \mathit{cl}_q(e^\prime) = c^\prime \\
							& \mbox{\scriptsize (\theequation.4)}	& \wedge \; \mathit{con}_q(e) = \tup{\wc{}, \wc{}, t} \\
							& \mbox{\scriptsize (\theequation.5)}	& \wedge \; \mathit{con}_q(e^\prime) = \tup{\wc{}, \wc{}, t^\prime} \\
							& \mbox{\scriptsize (\theequation.6)}	& \wedge \; p \in \mathit{allow}(t, t^\prime, c^\prime)
\end{array}
\label{eq:def_selxacf}
\end{equation}

\textbf{(Step 2.2)} 
In the next step, we identify those clauses in $\phi_\mathit{acf}$ that relate to common dynamic model components in the core-based sense. In \SELX{}, three such cases can be found:
\begin{enumerate}
	\item Clause (\ref{eq:def_selxacf}.1) checks for the presence of entities.
	\item Clauses (\ref{eq:def_selxacf}.2)--(\ref{eq:def_selxacf}.3) check for a specific class assigned to an entity.
	\item Clauses (\ref{eq:def_selxacf}.4)--(\ref{eq:def_selxacf}.5) check for a specific type assigned to an entity. The wildcards tell us that both user- and role-attributes of the security context are irrelevant for this comparison.
\end{enumerate}
We can ignore clause (\ref{eq:def_selxacf}.6), since no dynamic state change may influence the \textit{allow}-mapping.

Case 1 trivially implies that any new entity contributes to satisfying the ACF, which means that \emph{entity leaks} (effected by changing $E$) can be considered a first kind of unsafety. Similarly, case 2 hints at \emph{class assignment leaks} (via $\mathit{cl}$) and case 3 at \emph{type assignment leaks} (via $\mathit{con}$). We therefore identify three safety definitions for \SELX{}, based on these leakage targets $e$, $c$, and $t$ as follows:

\begin{definition}
Given a \SELX{} model $\tup{Q, \Sigma, \delta, \lambda, q_0, \Ext }$, a state $q \in Q$ is {\bf (e)-unsafe} with respect to an entity $e \in E$ iff $\exists\, q^\prime \in \{ \delta^\ast(q, a) \,|\, a \in \Sigma^\ast \}$:
\begin{equation*}
e \in E_{q^\prime} \wedge e \notin E_q \, .
\end{equation*}
\label{def:e-unsafe}
\vspace*{-2em}
\end{definition}

\begin{definition}
Given a \SELX{} model $\tup{Q, \Sigma, \delta, \lambda, q_0, \Ext }$, a state $q \in Q$ is {\bf (c)-unsafe} with respect to an object class $c \in C$ iff $\exists\, q^\prime \in \{ \delta^\ast(q, a) \,|\, a \in \Sigma^\ast \}$:
\begin{equation*}
\exists e \in E_{q^\prime} \cap E_q : \mathit{cl}_{q^\prime}(e) = c \wedge \mathit{cl}_q(e) \neq c \, .
\end{equation*}
\label{def:c-unsafe}
\vspace*{-2em}
\end{definition}

\begin{definition}
Given a \SELX{} model $\tup{Q, \Sigma, \delta, \lambda, q_0, \Ext }$, a state $q \in Q$ is {\bf (t)-unsafe} with respect to a type $t \in T$ iff $\exists\, q^\prime \in \{ \delta^\ast(q, a) \,|\, a \in \Sigma^\ast \}$:
\begin{equation*}
\begin{array}{l}
\exists e \in E_{q^\prime} \cap E_q : 
\mathit{con}_{q^\prime}(e) = \tup{\wc{}, \wc{}, t}
\wedge	 \mathit{con}_q(e) = \tup{\wc{}, \wc{}, t_q}
\wedge	t \neq t_q \, .
\end{array}
\end{equation*}
\label{def:t-unsafe}
\vspace*{-2em}
\end{definition}

\textbf{(Step 2.3)} 
Two conclusions can be drawn from the three definitions above: first, irrelevant safety definitions can be filtered to some extent by leveraging the information provided by core- and EL-semantics. In case of Def.~\ref{def:c-unsafe}, it becomes clear by studying \POST{} expressions in the state transition scheme (Fig.~\ref{fig:selx_bcmd-defs}) that entity classification is only performed during entity creation. Alternatively, the same fact can be inferred from the absence of any reference to the label set $C$ within $\SELX{}|_\mathrm{RR}$. 

Second, these safety definitions are deliberately demarcated with respect to their targets, which means they can be combined in a straight-forward way. As an example, an alternate definition of \emph{(t)-safety} could also cover any new entity not present in $q$. This results in a definition of \emph{(t)-simple-safety} aligned with Def.~\ref{def:r-simple-unsafe}:

\begin{definition}
Given a \SELX{} model $\tup{Q, \Sigma, \delta, \lambda, q_0, \Ext }$, a state $q \in Q$ is {\bf (t)-simple-unsafe} with respect to a type $t \in T$ iff $\exists\, q^\prime \in \{ \delta^\ast(q, a) \,|\, a \in \Sigma^\ast \}$:
\begin{equation*}
\begin{array}{rl}
& q \mbox{ is (t)-unsafe with respect to } t \\
\vee & \big( \exists e \in E_{q^\prime} : 
       \mathit{con}_{q^\prime}(e) = \tup{\wc{}, \wc{}, t} 
       \wedge q \mbox{ is (e)-unsafe with respect to } e \big) \, .
\end{array}
\end{equation*}
\label{def:t-simple-unsafe}
\vspace*{-1em}
\end{definition}

Fixing independent variables such as $e$ in Def.~\ref{def:t-unsafe} opens up even more variants of safety, similar to the specializations of \emph{(r)-simple-safety} presented in \cite{Tripunitara13a}. For example, a specialized definition of \emph{(e,t)-safety} could target a combination of both an entity leak and a type assignment leak. This task of refining safety definitions is, as above, subject to human interaction that provides additional semantic information about relevant analysis goals.

\subsection{\fds{} for \SELX{}}
\label{sec:eval-heuristic}

We now tailor \fds{} to \emph{(t)-safety} as an example. The safety definition is the foundation for determining heuristics behavior, which is defined through the model-specific modules introduced in Sec.~\ref{sec:fds}. The tailoring of these modules is implemented in the algorithms \ref{alg:selx-isLeaked}--\ref{alg:selx-createCDGSink}, given in Appendix~\ref{sec:appendix}. For the scope of this paper, we have opted for omitting a detailed discussion of \emph{assignParams}, which is still subject to ongoing work.

Firstly, the function \SELX{}::tUnsafe::isLeaked (Alg.~\ref{alg:selx-isLeaked}) is a direct implementation of the unsafety property given in Def.~\ref{def:t-unsafe}. 
The actual dependency analysis is performed by \SELX{}::tUnsafe::buildPredSet (Alg.~\ref{alg:selx-buildPredSet}). Based on some CDG vertex $c = \tup{\mathit{succ}, x_\mathit{succ}, \PRE{}(\mathit{succ}), \POST{}(\mathit{succ})}$, the algorithm builds a set
\begin{equation*}
\begin{array}{l}
\{ 
c_\mathit{pred} = \tup{\mathit{pred}, x_\mathit{pred}, \PRE{}(\mathit{pred}), \POST{}(\mathit{pred})} \in \Delta \,|\, \exists t \in T: \\
\PRE{}(\mathit{succ}) \mbox{ and } \POST{}(\mathit{pred}) \mbox{ depend on the value of } t 
\}
\end{array}
\end{equation*}
of commands whose prior execution $c$ depends on. In the actual implementation, label assignment model components ($\SELX{}|_\mathrm{LA}$) in pre- and post-conditions are checked to identify dependencies. An auxiliary set $\mathit{TDep}$ is then used to store and compare types these dependencies relate to.

\textit{createCDGSource}, implemented as \SELX{}::createCDGSource (Alg.~\ref{alg:selx-createCDGSource}), can be inferred easily if multi-level command specification is used to define a distinguished \emph{create} command: in that case, $\POST{}(\mathit{cmd})$ in Alg.~\ref{alg:selx-createCDGSource} matches $\POST{}(\mathit{create})$ in the state transition scheme.

At last, the implementation of \textit{createCDGSink} as \SELX{}::tUnsafe::createCDGSink (Alg.~\ref{alg:selx-createCDGSink}) must be a direct consequence from the implementation of isLeaked.
  \section{Conclusions}
\label{sec:concl}

This paper aimed at a uniform formal framework, tailorable for dynamic analysis of a wide range of access control models. The high-level goal behind this work is to demonstrate the practical benefits of combining semantically enhanced modeling patterns, such as EL and core-based modeling, when it comes to model analysis in a holistic security engineering process.

To achieve this, we have adapted the \ds{} heuristic safety analysis algorithm to an access control model for the SELinux operating system. Model-independent interfaces have been introduced into the heuristic, which are then tailored to the SELinux model in a workflow that is guided by semantic knowledge drawn from the modeling patterns. Based on such guiding rules, this workflow is to a significant degree automatable. Because the main portion of human intelligence is required in the early systems design phase of requirements engineering, we argue that this approach reduces later design and verification errors introduced through manually deriving formal methods for these steps.

Subject to future work is a thoroughly formalized, rule-based framework for heuristic tailoring, which enables a prototypical implementation of a model engineering toolkit to automate the process as far as possible. We have also started to evaluate the \fds{} heuristic for \SELX{} in terms of efficiency and success, which is subject to further improvements that aim at analyzing real-world systems in a productive setting.

\clearpage

\appendix
  \section{Appendix}
\label{sec:appendix}

\begin{algorithm}[H]
\KwIn{
	$q = \tup{E, \mathit{cl}, \mathit{con}}$ \dots{} model state the safety of which is to be analyzed \newline
	$q^\prime = \tup{E^\prime, \mathit{cl}^\prime, \mathit{con}^\prime}$ \dots{} model state to check for a leak \newline
	$t_\mathit{target}$ \dots{} leakage target type
}
\KwOut{
	$\top$ iff $q$ is (t)-simple-unsafe w.\,r.\,t. $t_\mathit{target}$
}
\BlankLine
\For{$e \in E^\prime \cap E$}{
  $\tup{\wc{}, \wc{}, t} \leftarrow \mathit{con}(e)$\;
  $\tup{\wc{}, \wc{}, t^\prime} \leftarrow \mathit{con}^\prime(e)$\;
  \lIf{$t^\prime = t_\mathit{target} \wedge t \neq t_\mathit{target}$}{
    \Return{$\top$}
  }
}
\Return{$\bot$}\;
\caption{\SELX{}::tUnsafe::isLeaked}
\label{alg:selx-isLeaked}
\end{algorithm}

\begin{algorithm}[H]
\KwIn{
	$\Delta$ \dots{} model's state transition scheme \newline
	$c = \tup{\mathit{succ}, x_\mathit{succ}, \PRE{}(\mathit{succ}), \POST{}(\mathit{succ})}$ \dots{} command whose predecessors are to be found
}
\KwOut{
	$P \subseteq \Delta$ \dots{} set of predecessor commands to $c$
}
\BlankLine
$P \leftarrow \emptyset$\;
$\mathit{TDep} \leftarrow \emptyset$\;
\For{$\phi \in \{ \phi_1 \dots \phi_n \,|\, \PRE{}(\mathit{succ}) = \logexpr{ \underline{\phi_1} \wedge \dots
 \wedge \underline{\phi_n} } \}$}{
  \lIf{$\phi = \logexpr{\mathit{con}_q(\wc{}) = \tup{\wc{}, \wc{}, \underline{t}}}$}{
    $\mathit{TDep} \leftarrow \mathit{TDep} \cup \{ t \}$
  }
}
\For{$c_\mathit{pred} = \tup{\mathit{pred}, x_\mathit{pred}, \PRE{}(\mathit{pred}), \POST{}(\mathit{pred})} \in \Delta$}{
  \For{$\psi \in \{ \psi_1 \dots \psi_n \,|\, \POST{}(\mathit{pred}) = \logexpr{ \underline{\psi_1} \wedge \dots
 \wedge \underline{\psi_n} } \}$}{
    \If{$\psi = \logexpr{\mathit{con}_{q^\prime} = \mathit{con}_q[\wc{} \mapsto \tup{\wc{}, \wc{}, \underline{t}}]}
         \wedge t \in \mathit{TDep}$}{
      $P \leftarrow P \cup \{ c_\mathit{pred} \}$\;
    }
  }
}
\Return{$P$}\;
\caption{\SELX{}::tUnsafe::buildPredSet}
\label{alg:selx-buildPredSet}
\end{algorithm}

\begin{algorithm}[H]
\KwIn{
	$q = \tup{E, \mathit{cl}, \mathit{con}}$ \dots{} model state the safety of which is to be analyzed
}
\KwOut{
	$\mathit{cmd}$ \dots{} identifier for the virtual CDG source command \newline
	$x_\mathit{cmd}$ \dots{} sequence of formal parameters for $\mathit{cmd}$ \newline
	$\PRE{}(\mathit{cmd})$ \dots{} pre-condition of $\mathit{cmd}$ \newline
	$\POST{}(\mathit{cmd})$ \dots{} post-condition of $\mathit{cmd}$
}
\BlankLine
$\mathit{cmd} \leftarrow$ ``virtualSourceCmd''\;
$x_\mathit{cmd} \leftarrow \epsilon$\;
$\PRE{}(\mathit{cmd}) \leftarrow \logexpr{\top}$\;
$\POST{}(\mathit{cmd}) \leftarrow \logexpr{\top}$\;

\For{$e \in E$}{
  $\begin{array}{rcl}
  \psi_e &\leftarrow& \logexpr{E_{q^\prime} = E_q \cup \{ \underline{e} \}\\
    &&\wedge\, \mathit{cl}_{q^\prime} = \mathit{cl}_q[\underline{e} \mapsto \underline{\mathit{cl}(e)}]\\
    &&\wedge\, \mathit{con}_{q^\prime} = \mathit{con}_q[\underline{e} \mapsto \underline{\mathit{con}(e)}]} \mbox{\;}
  \end{array}$\newline
  $\POST{}(\mathit{cmd}) \leftarrow \logexpr{\underline{\POST{}(\mathit{cmd})} \wedge \underline{\psi_e}}$\;
}
\Return{$\mathit{cmd}, x_\mathit{cmd}, \PRE{}_\mathit{cmd}, \POST{}_\mathit{cmd}$}\;
\caption{\SELX{}::createCDGSource}
\label{alg:selx-createCDGSource}
\end{algorithm}

\begin{algorithm}[tb]
\KwIn{
	$t_\mathit{target}$ \dots{} leakage target type
}
\KwOut{
	$\mathit{cmd}$ \dots{} identifier for the virtual CDG sink command \newline
	$x_\mathit{cmd} \in \Sigma_X$ \dots{} sequence of formal parameters for $\mathit{cmd}$ \newline
	$\PRE{}(\mathit{cmd})$ \dots{} pre-condition of $\mathit{cmd}$ \newline
	$\POST{}(\mathit{cmd})$ \dots{} post-condition of $\mathit{cmd}$
}
\BlankLine
$\mathit{cmd} \leftarrow$ ``virtualSinkCmd''\;
$x_\mathit{cmd} \leftarrow \epsilon$\;
$\PRE{}(\mathit{cmd}) \leftarrow \logexpr{\exists e \in E_q : \mathit{con}_q(e) = \tup{\wc{}, \wc{}, \underline{t_\mathit{target}}}}$\;
$\POST{}(\mathit{cmd}) \leftarrow \logexpr{\top}$\;
\Return{$\mathit{cmd}, x_\mathit{cmd}, \PRE{}_\mathit{cmd}, \POST{}_\mathit{cmd}$}\;
\caption{\SELX{}::tUnsafe::createCDGSink}
\label{alg:selx-createCDGSink}
\end{algorithm}

\bibliographystyle{splncs03}

\vfill
\end{document}